\documentclass[aps,prb,preprint,groupedaddress]{revtex4-1}
\usepackage[latin9]{inputenc}
\usepackage{graphicx}

\makeatletter
 
 \@ifundefined{textcolor}{}
 {%
   \definecolor{BLACK}{gray}{0}
   \definecolor{WHITE}{gray}{1}
   \definecolor{RED}{rgb}{1,0,0}
   \definecolor{GREEN}{rgb}{0,1,0}
   \definecolor{BLUE}{rgb}{0,0,1}
   \definecolor{CYAN}{cmyk}{1,0,0,0}
   \definecolor{MAGENTA}{cmyk}{0,1,0,0}
   \definecolor{YELLOW}{cmyk}{0,0,1,0}
 }

\usepackage{dcolumn}
\usepackage{bm}

\makeatother

\begin{document}






\title{Theoretical polarization dependence of the two-phonon double-resonant
Raman spectra of graphene}

\author{Valentin N. Popov}

\affiliation{Faculty of Physics, University of Sofia, BG-1164 Sofia, Bulgaria}

\author{Philippe Lambin}

\affiliation{Research Center in Physics of Matter and Radiation, Facult{\'e}s Universitaires
Notre Dame de la Paix, B-5000 Namur, Belgium}

\date{\today}
\begin{abstract}
The experimental Raman spectra of graphene exhibit a few intense two-phonon
bands, which are enhanced through double-resonant scattering processes.
Though there are many theoretical papers on this topic, none of them
predicts the spectra within a single model. Here, we present results
for the two-phonon Raman spectra of graphene calculated by means of
the quantum perturbation theory. The electron and phonon dispersions,
electronic lifetime, electron-photon and electron-phonon matrix elements,
are all obtained within a density-functional-theory-based non-orthogonal
tight-binding model. We study systematically the overtone and combination
two-phonon Raman bands, and, in particular, the energy and polarization
dependence of their Raman shift and intensity. We find that the ratio
of the integrated intensities for parallel and cross polarized light
for all two-phonon bands is between $0.33$ and $0.42$. Our results
are in good agreement with the available experimental data. 
\end{abstract}
\maketitle


\section{Introduction}

Presently, graphene is considered as a prospective material for nanoelectronics
and nanophotonics.\cite{geim07,bona10} Among the various experimental
techniques, the Raman spectroscopy has proven to be an indispensable
tool for investigation of this material.\cite{ferr06,mala09a,dres10,zoly11}

Graphene has a single Raman-active phonon $E_{2g}$ observed as an
intense line (the G band) in the first-order Raman spectra. The second-order
spectra of graphene with low defect density has several intense bands,
which originate from scattering of electrons and holes by two phonons
of the same/different frequency and non-zero momentum and are called
overtone/combination bands. The appearance of intense second-order
bands can be explained by the double-resonant (DR) scattering mechanism.\cite{thom00,reic04}
These bands contain valuable information on the phonon dispersion\cite{sait02,grun02,mafr07,grun09}
and the electron-phonon and electron-electron matrix elements.\cite{bask09}
The Raman spectra of graphene with defects show additional bands,
which arise from DR scattering of electrons and holes by phonons and
defects.

The theoretical investigation of the two-phonon DR scattering in graphene
has been performed using various approximations: replacement of the
electron-photon and electron-phonon interactions with constants, using
a constant electronic lifetime, considering only high-symmetry directions
in the Brillouin zone, as well as exact DR conditions. The predicted
dispersive behavior of the Raman bands\cite{thom00} and the frequency 
shift of the Stokes and anti-Stokes Raman bands\cite{canc02} have 
been found in quantitative agreement with the experimental data. It 
has also been realized that the integration over the entire Brillouin 
zone of graphene is essential for predicting the Raman intensity.\cite{maul04a,naru08} 
While most of the theoretical papers focus on the most intense overtone 
band, in a recent study, all two-phonon Raman bands with observable intensity have been calculated using an electron-phonon matrix element derived within a nearest-neighbor $\pi$-band tight-binding model.\cite{vene11} 
The dominant contribution to the two-phonon bands from different parts of the 
Brillouin zone\cite{vene11,naru11} and from different scattering processes\cite{vene11} has also been discussed. The polarization dependence of the most intense
overtone band has been studied experimentally and theoretically.\cite{yoon08}
The progress, made so far in the modeling of the DR bands, has been
achieved either with simple $\pi$-band tight-binding models, or with
more sophisticated models but relying on approximations of the 
electron-phonon matrix element, and
in most cases concerns only a few intense bands. Our experience
indicates that, although the Raman shift of the bands is not sensitive
to the used matrix element, their Raman intensity crucially depends
on it.

Here, we calculate the two-phonon DR Raman spectra of graphene using
a non-orthogonal tight-binding (NTB) model, which implements parameters
derived from a density functional theory (DFT) study and thus has no adjustable parameters.\cite{pore95}
In particular, the electronic\cite{popo04} and phonon\cite{popo06}
dispersion, the electron-photon and electron-phonon matrix elements,\cite{popo05}
as well as the electronic linewidth\cite{popo06a} are all obtained
within this model. The NTB model for electrons and phonons in graphene
is introduced in Sec. II. The calculated two-phonon DR Raman spectra
of graphene and its polarization dependence are discussed in Sec.
III. The paper ends up with conclusions (Sec. IV).

\section{Theoretical part}

\subsection{The NTB model}

We use a NTB model with four valence electrons per carbon atom to
calculate the electronic dispersion of graphene.\cite{popo04} This
model is based on matrix elements of the Hamiltonian and overlap matrix
elements derived from DFT\cite{pore95} and therefore
it does not rely on any adjustable parameters. It also allows one
to estimate the total energy and the forces on the atoms. This feature
is utilized for relaxation of the atomic structure. Up to a few electron
volts away from the Fermi energy, the electronic structure of graphene
has the form of conic valence and conduction bands (Dirac cones) with
a common apex (the Dirac point) at two non-equivalent special points,
K and K$^{'}$, of the Brillouin zone. This specific form of the electronic
bands plays an important role in the enhancement of the two-phonon
Raman scattering through the DR mechanism.

The dynamical model of graphene uses a dynamical matrix derived by
a perturbative approach within the NTB model.\cite{popo06} The electron-photon
and electron-phonon matrix elements are calculated explicitly.\cite{popo05}
The summation over the Brillouin zone in the first-order perturbation
term of the dynamical matrix is performed over a $40\times40$ mesh
of $\mathbf{k}$ points, for which the phonon frequencies converge 
within $1$ cm$^{-1}$. The calculated in-plane phonon branches
of graphene, after scaling by a factor of $0.9$, agree fairly
well with the available experimental data\cite{popo06} (Fig. 1).
The phonons with displacement in the graphene plane (in-plane phonons)
can interact with electrons and thus can contribute to the Raman spectra.
Those with atomic displacement perpendicular to graphene (out-of-plane
phonons) are less well reproduced but they do not contribute to the
spectra.

It will be shown below that only phonons, close to the high-symmetry
directions $\Gamma$K, $\Gamma$M, and KM of the Brillouin zone, are
of major importance for the two-phonon spectra. The phonon branches
along these directions will be denoted, as usual, by two-letter acronyms describing
their vibrational pattern: the letters O and A stand for ``optical''
and ``acoustic'', respectively; the letters L, T, and Z denote in-plane
longitudinal, in-plane transverse, and out-of-plane atomic displacement,
respectively. The acronyms for the branches along the KM direction
will be primed. Alternatively, for each wavevector, the phonons with
be ascribed the index $\nu$, $\nu=1,...,6$, in order of increasing
frequency. The phonons with a certain $\nu$ can belong to branches
with a different vibrational pattern. For example, a phonon with $\nu=6$
can belong to the LO or TO branch. It will also be argued that only
phonons close to the $\Gamma$ and K points give a significant contribution
to the two-phonon Raman spectra. Such phonons will be denoted by acronyms
ending with @$\Gamma$ and @K. For example, LO phonons close to the
$\Gamma$ point will be denoted by LO@$\Gamma$ and TO phonons close
to the K point along the KM direction will be denoted by TO$^{'}$@K
(see, Fig. 1).

\subsection{The double-resonant processes}

The amplitude for two-phonon DR Raman scattering processes in graphene
is described by fourth-order terms in perturbation theory.\cite{mart83}
The underlying processes include virtual scattering of electrons/holes by phonons
between states of the Dirac cone at the K point or the K$^{'}$ point,
or between states of the Dirac cones at the K and K$^{'}$ points.
The momentum is conserved in each virtual process but the energy is
conserved only for the entire DR process. Below, we will consider
only Stokes processes. In this case, a two-phonon DR process includes
an absorption of a photon with a creation of an electron-hole pair,
two consecutive processes of scattering of an electron/hole with creation
of a phonon, and a recombination of the electron-hole pair with
an emission of a photon (Fig. 2). There are altogether eight such
processes.\cite{kurt02,vene11} The total two-phonon Raman intensity
is given by the expression

\begin{equation}
I\propto\sum_{f}\left|\sum_{c,b,a}\frac{M_{fc}M_{cb}M_{ba}M_{ai}}{\left(E_{i}-E_{c}-i\gamma\right)\left(E_{i}-E_{b}-i\gamma\right)\left(E_{i}-E_{a}-i\gamma\right)}\right|^{2}\delta\left(E_{i}-E_{f}\right)\label{a2}
\end{equation}

Here, the inner sum is the scattering amplitude. $E_{u}$, $u=i,a,b,c,f$,
are the energies of the initial ($i$), intermediate ($a,b,c$), and
final ($f$) states of the system of photons, electrons, holes, and
phonons. In the initial state, only an incident photon is present
and, therefore, $E_{i}=E_{L}$, where $E_{L}$ is the incident photon
energy. In the final state, there is a scattered photon and two created
phonons. $M_{uv}$ are the matrix elements for virtual processes between
initial, intermediate, and final states. In particular, $M_{ai}$
and $M_{fc}$ are the matrix elements of momentum for the processes
of creation and recombination of an electron-hole pair, respectively.
$M_{ba}$ and $M_{cb}$ are the electron/hole-phonon matrix elements.
$\gamma$ is the sum of the halfwidths of pairs of electronic and
hole states, and will be referred to as the electronic linewidth.
The electron-photon and electron-phonon matrix elements, and the electronic
linewidth are calculated explicitly.\cite{popo05,popo06a} The Dirac
delta function ensures energy conservation for the entire process.
In the calculations, it is replaced by a Lorentzian with a halfwidth
of $5$ cm$^{-1}$. The summation over the intermediate states runs
over all valence and conduction bands, and over all electron wavevectors
$\mathbf{k}$. The summation over the final states runs over all phonon
branches and phonon wavevectors $\mathbf{q}$. For both summations,
convergence is reached with a $800\times800$ mesh of $\mathbf{k}$
and $\mathbf{q}$ points in the Brillouin zone.

For the discussion of the polarization dependence of the Raman intensity
of the two-phonon bands, it is advantageous to rewrite Eq. (\ref{a2})
in the form

\begin{equation}
I\propto\sum_{f}\left|\mathbf{e}_{S}\cdot R\cdot\mathbf{e}_{L}\right|^{2}\delta\left(E_{i}-E_{f}\right)\label{a4}
\end{equation}
 Here, $\mathbf{e}_{L}$ and $\mathbf{e}_{S}$ are the polarization
vectors of the incident and scattered laser light, respectively, and $R$ is the
Raman tensor. Everywhere below we consider only backscattering geometry
in accord with the usual experimental Raman setup for graphene and,
therefore, the polarization vectors lie in the graphene plane.

\section{Results and Discussion}

\subsection{Electronic linewidth}

The electronic linewidth $\gamma$ is due to a large extent to scattering
of the electrons (holes) by phonons and other electrons (holes). The
majority of the published reports assume that $\gamma$ is energy-independent.
Recently, it has been argued that in undoped graphene $\gamma$ is
dominated by electron-phonon processes and the expression $\gamma=9.44E+3.40E^{2}$
has been derived, where the energy separation between the valence
and conduction bands $E$ is in eV and $\gamma$ is in meV.\cite{vene11}
Here, the electronic linewidth was calculated by summing up the contributions
of all electron/hole-phonon scattering processes for all phonons in
the Brillouin zone as a function of $E$. The obtained energy dependence
was approximated in the range $[1.0,3.5]$ eV with the expression
\begin{equation}
\gamma=12.60E+3.45E^{2}\label{b2}
\end{equation}
 For energies in this energy range, $\gamma$ changes more than four
times from $16$ to $86$ meV (Fig. 3).

\subsection{Overtone bands}

The calculated overtone Raman spectrum for $E_{L}=2.0$ eV and parallel light polarization along a zigzag line of carbon bonds  is shown in
Fig. 4 (bottom). It has two intense bands ($2D$ and $2D^{'}$) and
three weaker ones ($2D^{3}$, $2D^{4}$, and $2D^{''}$) and a much
weaker one ($2D^{5}$). There are also other bands of in-plane phonons
close to the $\Gamma$ and K points but they are either very weak,
or are in the shoulders of intense bands, and in both cases are practically
unobservable.

The assignment of the overtone spectrum can be performed by analyzing
the contribution of phonons with different $\nu$ and from different
parts of the Brillouin zone of graphene. First, the contributions
of the phonons with $\nu=2,...,6$ are given in Fig. 4 (top five graphs).
The bands in these spectra originate from pairs of phonons TA@$\Gamma$
($\nu=2$), LA@$\Gamma$ and TA@K ($\nu=3$), LA@K ($\nu=4$), TO@K
($\nu=5$), and TO$^{'}$@K and LO@K ($\nu=6$) (see also Fig. 1).
The contributions for $\nu=5,6$ are by about three orders of magnitude
larger than the remaining ones.

Secondly, the assignment of the Raman bands to definite phonons at
the $\Gamma$ and K points is supported by the analysis of the contributions
to the spectra from different parts of the Brillouin zone for parallel
light polarization with averaging over all orientations in the graphene
plane. In Fig. 5, the regions with major contribution to the bands
are given by shaded areas. In particular, the $2D^{3}$ band comes
from phonons TA@$\Gamma$ along the $\Gamma$K direction, the $2D^{4}$
band - from phonons LA@$\Gamma$ along $\Gamma$M, the $2D^{''}$
band - from phonons LA@K along $\Gamma$K (not shown in Fig. 5), and the $2D^{'}$
band - from phonons LO@$\Gamma$ from all directions in the Brillouin
zone. The $2D$ band originates from phonons TO@K and TO$^{'}$@K, which contribute to 
the integrated Raman intensity of this band in ratio $\approx10:1$. It has long been accepted that the dominant contribution to
the $2D$ band comes from phonons along the KM direction and that
the contribution along the $\Gamma$K direction vanishes because of
interference.\cite{naru08} Our graphs in Fig. 4 and 5 for $\nu=5,6$  
confirm 
the recent conclusion that the main contribution to this band comes
from the $\Gamma$K direction and only a small part comes from the
KM direction.\cite{vene11,naru11} We also confirm the result of 
Ref. \cite{vene11}, that among the eight DR processes, those with scattering
of an electron and a hole have dominant contribution to the bands.

The overtone bands are dispersive, i.e., their Raman shift depends
on the laser energy (Fig. 6). The shift increases (decreases) due to
increase (decrease) of the phonon frequency away from the $\Gamma$
and K points. It is approximately linear in $E_{L}$ with slopes
of $\approx0$ for the $2D^{'}$ band, $89$ cm$^{-1}$/eV 
($2D$ band), $-193$ cm$^{-1}$/eV ($2D^{''}$),
$419$ cm$^{-1}$/eV ($2D^{4}$), and $256$ cm$^{-1}$/eV ($2D^{3}$) at $E_{L}=2.0$ eV .
The slope of the $2D$ band corresponds to the experimental one of $88$ cm$^{-1}$/eV
(Ref. {[}\cite{mafr07}{]}). 

The integrated Raman intensity of the overtone bands (Fig. 7) is quasi-linear
in $E_{L}$ except for the $2D^{3}$ band. The ratio of the integrated
intensities $A\left(2D\right)/A\left(2D^{'}\right)$ has been discussed
a lot in the literature because it is related to the electron-phonon matrix elements
at the K and $\Gamma$ points. Here, we find that this ratio depends
on the energy: it varies from $4.7$ for $E_{L}=1.0$ eV to $15.0$
for $E_{L}=3.5$ eV; for $E_{L}=2.4$ eV it is equal to $8.7$. 
By contrast, recent sophisticated DFT-GW calculations\cite{vene11}
yield $21.5$, which is several times larger than our value. The reason
for this disagreement can be found in the use of GW corrected electron-phonon
matrix elements for the LO phonon at the $\Gamma$ point and
the TO phonon at the K point. While the ratio of
the squares of these matrix elements at the two points derived by
DFT is $M_{K}^{2}/M_{\Gamma}^{2}=2.02$, the DFT-GW result is $3.03$
(Ref. {[}\cite{lazz08}{]}). Since the Raman intensity depends on
the square of this ratio, the DFT-GW gives an intensity ratio that
is larger than the DFT one by a factor of $2.25$. Our result, corrected by the same correction factor, is $19.6$. Both theoretical results
underestimate the experimental values of $26$ (Ref.
{[}\cite{alzi10}{]}) and $27$ (Ref. {[}\cite{ferr06}{]}) by $\sim25\%$.
The origin of this underestimation is still unknown.

\subsection{Combination bands}

The combination Raman spectrum is calculated for $E_{L}=2.0$ eV and parallel light polarization along a zigzag line of carbon bonds (Fig. 8, bottom).
The spectrum is dominated by two intense bands: $D+D^{''}$ and a higher
frequency one, which originate from the branches $5+4$ and $6+5$,
respectively. There are also four other less intense bands: $D^{'}+D^{3}$,
$D^{'}+D^{4}$, and $D+D^{5}$.

As above, the assignment is facilitated by considering the contributions
to these bands from pairs of phonons with $\nu,\nu^{'}=2,3,4,5,6$.
It is seen in Fig. 8 (top six graphs) that for each pair $\nu+\nu^{'}$
there is a single band. These bands are due to pairs of phonons TOTA@K
($\nu+\nu^{'}=5+3$), TOLA@K ($5+4$), LOTA@$\Gamma$ ($6+2$), LOLA@$\Gamma$
($6+3$), TO$^{'}$LO$^{'}$@K ($6+4$), and TO$^{'}$LA$^{'}$@K
($6+5$) (see also Fig. 1). Unlike the case of the overtone bands,
the intensity of the various combination bands varies only by one
order of magnitude. The band $D+D^{''}$ is mainly due to phonons
TOLA@K with a small contribution of phonons TO$^{'}$LO$^{'}$@K.
The phonons TO$^{'}$LA$^{'}$@K give rise to an intense band at $\sim2700$
cm$^{-1}$. The large contribution of the phonons TO$^{'}$LO$^{'}$@K
and TO$^{'}$LA$^{'}$@K is consistent with the large electron-phonon
matrix element for the phonon TO$^{'}$@K and the nonzero matrix element
for the phonons LA$^{'}$@K and LO$^{'}$@K. We note that the latter
contrubite to the overtone spectra as well but their bands are masked
in the shoulder of the intense $2D$ band. The band at $2700$ cm$^{-1}$
also overlaps considerably with the much more intense $2D$ band and
cannot be observed as a separate feature. Thus, the total Raman spectrum
exhibits the intense $2D$ band, the two weak bands $2D^{'}$ and
$D+D^{''}$(Fig. 9), as well as the two very weak bands $D^{'}+D^{3}$
and $D^{'}+D^{4}$.

Similarly to the overtone bands, the combination bands are dispersive.
The dependence of the Raman shift on the laser energy (Fig. 10) is
almost linear with a slope of $74$ cm$^{-1}$/eV for the highest-frequency
band, $-50$ cm$^{-1}$/eV ($D+D^{''}$), $-54$ cm$^{-1}$/eV ($D+D^{5}$),
$210$ cm$^{-1}$/eV ($D^{'}+D^{4}$), and $123$ cm$^{-1}$/eV ($D^{'}+D^{3}$)
at $E_{L}=2.0$ eV. Each slope is equal approximately to the sum of
half of the slopes of the corresponding overtone bands. The calculated
energy dependence of the $D+D^{''}$ band frequency reproduces fairly
well the experimental data but the derived slope is almost three times
larger than the experimental one of $-18$ cm$^{-1}$/eV for energies
between $1.92$ and $2.71$ eV (Ref. {[}\cite{mafr07}{]}). The calculated
frequencies for the $D^{'}+D^{4}$ and $D^{'}+D^{3}$ bands underestimate
the Raman data up to $~80$ and $~30$ cm$^{-1}$, respectively. Similar
underestimation is present in other precise calculations.\cite{vene11,sato11}
On the other hand, the frequency slopes of the  bands  $D^{'}+D^{4}$ and $D^{'}+D^{3}$
are in fair agreement with the measured ones of $221$ cm$^{-1}$/eV
and $140$ cm$^{-1}$/eV, respectively.\cite{cong11} We note that
neither we, nor the authors of Ref. {[}\cite{vene11}{]}, have found any observable
combination band for phonons TOLA@$\Gamma$ though such band has been
established\cite{cong11} by fitting a low-intensity Raman band with
two Lorentzians for two overlapping bands assigned to phonons TOLA@$\Gamma$
and LOLA@$\Gamma$.

The calculated integrated Raman intensity is shown in Fig. 11. The
curves are increasing functions of laser energy for all bands except
for the band $D^{'}+D^{3}$ and the highest frequency one. For the
most studied $D+D^{''}$ band, we find $A\left(2D\right)/A\left(D+D^{''}\right)=12.2$. Introducing the DFT-GW correction factor for the electron-phonon matrix element of the TO phonon at the K point of $1.76$ (Ref. \cite{lazz08}), for the latter ratio we obtain the value of $21.5$, which agrees with the previous estimate\cite{vene11} of $16$ and the experimental one\cite{ferr06} of $\approx21$. The inset of Fig. 11 shows the dependence of the DFT-GW corrected ratios $A\left(2D'\right)/A\left(2D\right)$ and 
$A\left(D+D^{''}\right)/A\left(2D\right)$ as a function of $E_{L}$. Both curves agree well with the theoretical ones\cite{vene11} and the available experimental data.\cite{ferr06}   

\subsection{Dependence on the electronic linewidth}

Most of the previously reported two-phonon Raman spectra have been
calculated for energy-independent $\gamma$. In order to study the
effect of this approximation we performed calculation of the integrated
intensity $A_{c}$ for constant linewidth $\gamma_{c}=\gamma(E_{L}=2.0$ eV$)=39$ meV as well.
We found that, for any of the two-phonon peaks discussed above, the
calculated ratio of the integrated intensities, $A$ and $A_{c}$,
for variable and constant linewidth, respectively, can be fitted very
well with the expression $A/A_{c}=\left(\gamma_{c}/\gamma\right)^{2}$.
Therefore, $A$ can be written as $A=f/\gamma^{2}$, where $f$ depends
on $E_{L}$ and does not depend on $\gamma$. It is clear from
Figs. 7 and 11 that the rate of change 
of $f$ as a function of $E_{L}$ is different for the different two-phonon bands. Thus, the
ratio of the integrated intensities for any pair of two-phonon bands $1$ and
$2$, $A_{1}/A_{2}=f_{1}/f_{2}$, should depend on $E_{L}$
in agreement with the recent conclusions.\cite{vene11} We note that, in the latter paper, deviation from the $1/\gamma^{2}$ behavior has been found. The correct dependence should be established by comparison with experimental Raman data collected at more values of the laser energy.   

The intensity ratio for the $2D$ and $2D^{'}$ overtone bands 
can be obtained using the simple tight-binding analytical formulas:\cite{bask08} $A\left(2D\right)\propto2\left(\gamma_{K}/\gamma\right)^{2}$
and $A\left(2D^{'}\right)\propto\left(\gamma_{\Gamma}/\gamma\right)^{2}$.
Here, $\gamma=\gamma_{\Gamma}+\gamma_{K}$ is the total electronic
linewidth; $\gamma_{\Gamma,K}\propto E_{L}M_{\Gamma,K}^{2}/\omega_{\Gamma,K}$
are the linewidths due to the LO phonon at the $\Gamma$ point and
the TO phonon at the K point, $M_{\Gamma,K}^{2}$, and $\omega_{\Gamma,K}$
are the corresponding square of the electron-phonon matrix element
and the phonon frequency, respectively. The predicted ratio of the integrated
intensity of the two bands is a constant, equal to
$3$ (or $7.9$), evaluated with electronic linewidth from DFT (or
DFT-GW).\cite{lazz08} The latter values are about $7$ (or $2.7$) times smaller than those of the precise derivations at $E_{L}=2.4$ eV, presented here and in Ref.
{[}\cite{vene11}{]}. The disagreement for the intensity ratio can be sought in the approximations used in the theoretical scheme of Ref. {[}\cite{bask08}{]}.

\subsection{Polarization dependence of the bands}

Let us choose a coordinate system in the graphene plane with $x$
axis along a zigzag line of carbon bonds (and therefore the $y$ axis
is along an armchair line of carbon bonds), and set $\mathbf{e}_{L}=(\cos\alpha,\sin\alpha)$
and $\mathbf{e}_{S}=(\cos\beta,\sin\beta)$. In the case of parallel
and cross polarizations, the intensity is not angle-dependent because
of the high-symmetry of graphene. On the contrary, for fixed polarization
angle and variable analyzer angle, or vice versa, the intensity depends
only on the difference $\alpha-\beta$ and does not depend on the
sign of this difference. Therefore, the Raman intensity, Eq. (\ref{a4}), can be written as

\begin{equation}
I=\left(I_{\mid\mid}-I_{\perp}\right)\cos^{2}\left(\alpha-\beta\right)+I_{\perp}\label{e8}
\end{equation}
 The intensity for parallel and cross polarized light, $I_{\mid\mid}$
and $I_{\perp}$, can be obtained by fitting this expression to the
calculated angular dependence of the intensity. Expression, similar
to Eq. (\ref{e8}), holds for the integrated intensity as well.

Fig. 12 shows the polarization dependence of the ratio of the
calculated integrated intensities $A_{\perp}/A_{\mid\mid}$ at $E_{L}=2.0$
eV. As could be expected, all two-phonon bands
have similar angular dependence of the Raman intensity, which follows
from Eq. (\ref{e8}). The derived values of the ratio $A_{\perp}/A_{\mid\mid}$
are: $0.3678$ ($2D^{3}$), $0.4195$ ($2D^{4}$), $0.3321$ ($2D^{''}$),
$0.3377$ ($2D$), $0.4075$ ($2D^{'}$), $0.3718$ ($D^{'}+D^{3}$),
$0.4073$ ($D^{'}+D^{4}$), $0.3401$ ($D+D^{5}$), $0.3350$ ($D+D^{''}$).
The intensity ratio for all bands is between $0.33$ and $0.42$.
The ratio for the $2D$ band corresponds to the value $1/3$ derived
within a simple tight-binding model\cite{bask08} and the tight-binding
estimate\cite{yoon08} of $0.32$, and is in fair agreement with the
experimental value\cite{yoon08} of $0.3$. So far as we are aware,
there are no reports on the polarization dependence of the other two-phonon
bands.

\section{Conclusions}

We have presented a complete theoretical treatment of the two-phonon
Raman bands of graphene within a non-orthogonal tight-binding model
with no adjustable parameters and no other approximations. We have
calculated the laser energy and polarization dependence of the Raman
shift and intensity. In particular, the ratio of the integrated Raman
intensity for parallel and cross polarized light for all bands is
between $0.33$ and $0.42$. The agreement of our results with available
experimental data is very good and the predictions can be used for
further comparison to experiment and for support in the assignment
of the two-phonon Raman spectra. Furthermore, we have made definite
conclusions on the dominant contributions to the Raman scattering
amplitude. Our computations can easily be extended to include higher-order
Raman processes.

\acknowledgments

V.N.P. acknowledges financial support from Facult{\'e}s Universitaires
Notre Dame de la Paix, Namur, Belgium, and Grant No.71/05.04.2012 of University of Sofia, Sofia, Bulgaria.

%

\begin{figure}[tbph]
\includegraphics[width=80mm]{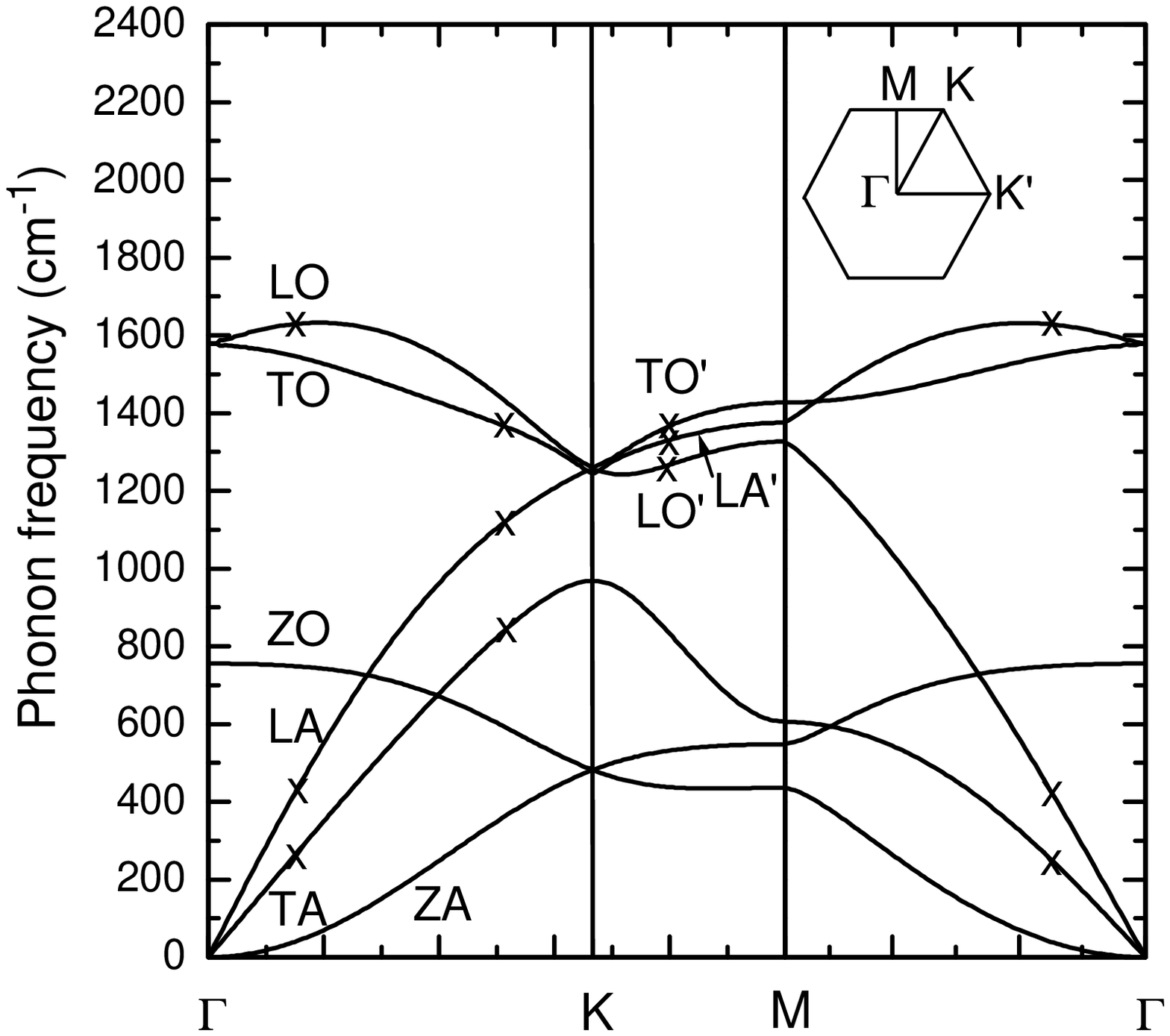} 
\caption{Calculated phonon dispersion of graphene along the high-symmetry directions
in the Brillouin zone.\cite{popo06} The six phonon branches are marked
by the acronyms LO, TO, ZO, LA, TA, and ZA. The letters O and A stand
for ``optical'' and ``acoustic'', respectively; L, T, and Z denote
in-plane longitudinal, in-plane transverse, and out-of-plane atomic
displacement, respectively. The crosses mark the phonons which play
major role in the DR processes. Inset: the hexagonal Brillouin zone
of graphene with the special points $\Gamma$, M, K, and K$^{'}$.}
\end{figure}
\begin{figure}[tbph]
\includegraphics[width=80mm]{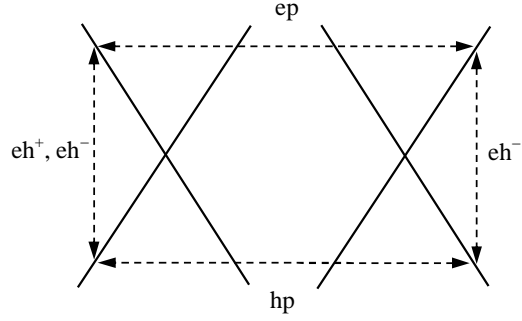} 
\caption{Schematic representation of the DR processes in graphene. The solid
lines are a cross-section of the Dirac cones at the K and K$^{'}$
points of the Brillouin zone. The dashed lines are virtual processes:
``eh$^{+}$'' and ``eh$^{-}$'' are electron (e) - hole (h) creation
and annihilation processes, respectively; ``ep'' and ``hp'' are
electron and hole scattering processes by a phonon (p). DR processes
can also take place between bands at the K or K$^{'}$ point only.}
\end{figure}
\begin{figure}[tbph]
\includegraphics[width=80mm]{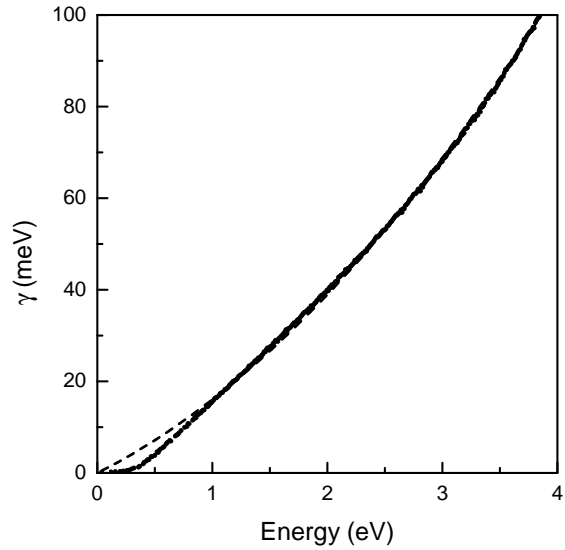} 
\caption{Calculated electronic linewidth $\gamma(E)$ (solid symbols). The
dashed line is a fit to the calculated data. The fitting function
is given by Eq. (\ref{b2}). }
\end{figure}
\begin{figure}[tbph]
\includegraphics[width=80mm]{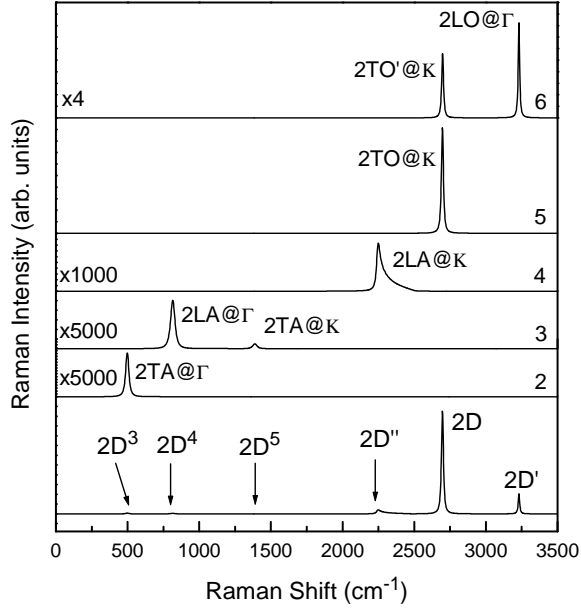} 
\caption{Calculated overtone Raman spectrum for $E_{L}=2.0$ eV (bottom). The
contributions from phonons with $\nu=2,...,6$ are shown in the top
graphs. The notation of the Raman bands $2D^{3}$, $2D^{4}$ and $2D^{5}$
is taken from Ref. {[}\cite{vene11}{]}.}
\end{figure}
\begin{figure}[tbph]
\includegraphics[width=80mm]{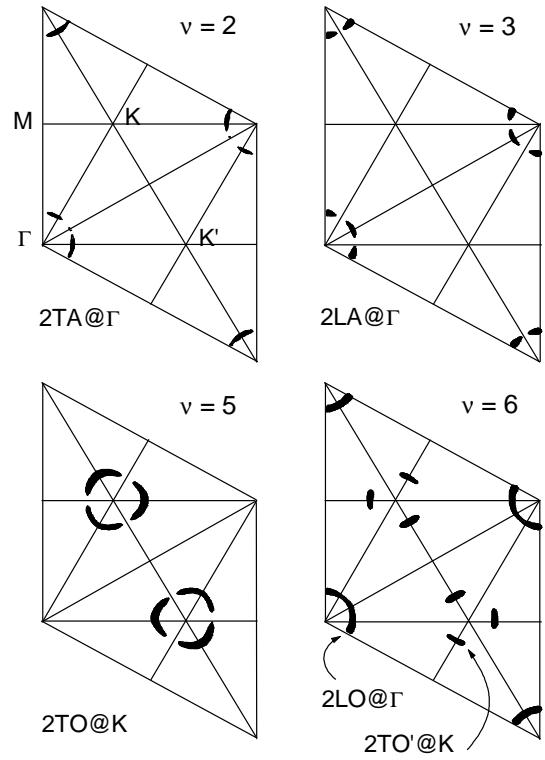} 
\caption{The contribution to the overtone bands from phonons with $\nu=2,3,5,6$
from different parts of the rhombic Brillouin zone of graphene. The
graph for $\nu=4$ for phonons 2LA@K (not shown) is similar to that
for $\nu=5$. }
\end{figure}
\begin{figure}[tbph]
\includegraphics[width=80mm]{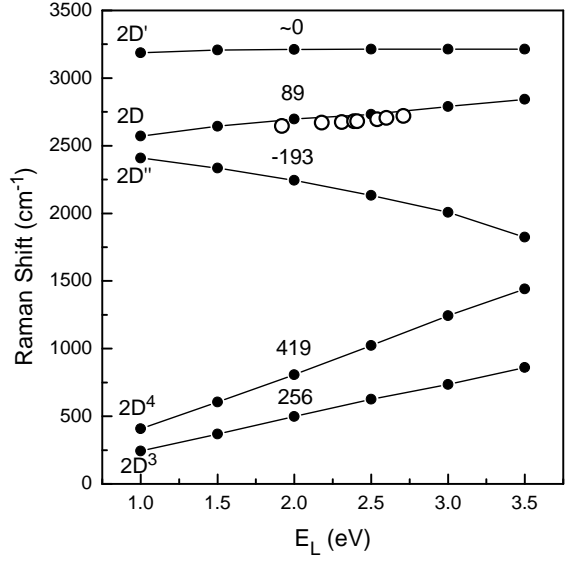} 
\caption{Calculated dependence of the Raman shift of the overtone bands on
$E_{L}$ (solid symbols). The numbers are the slopes of the curves
for $E_{L}=2.0$ eV. The empty symbols are experimental data.\cite{mafr07}}
\end{figure}
\begin{figure}[tbph]
\includegraphics[width=80mm]{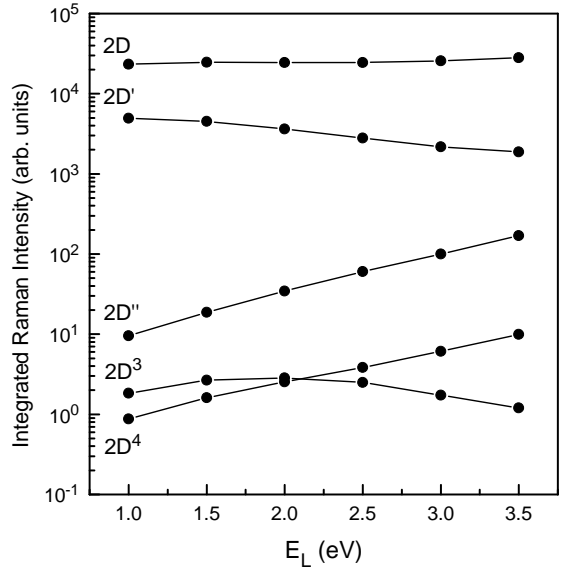} 
\caption{Calculated dependence of the integrated Raman intensity of the overtone
bands on $E_{L}$ (solid symbols).}
\end{figure}
\begin{figure}[tbph]
\includegraphics[width=80mm]{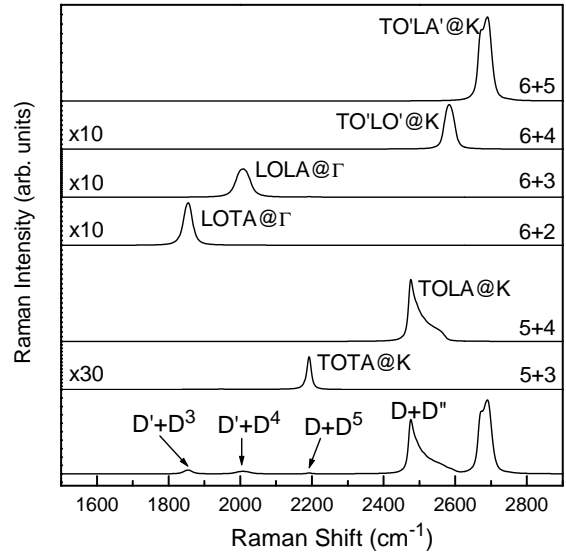} 
\caption{Calculated combination Raman spectrum for $E_{L}=2.0$ eV (bottom).
The contributions from pairs of phonons with $\nu+\nu'=5+3,...,6+5$
are shown in the top graphs. }
\end{figure}
\begin{figure}[tbph]
\includegraphics[width=80mm]{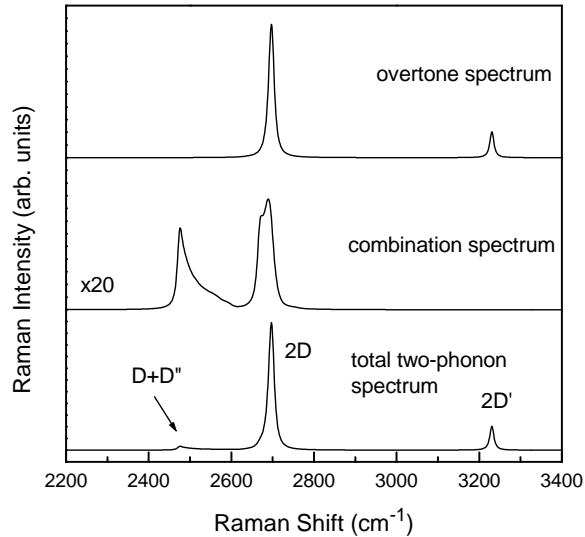} 
\caption{Calculated total two-phonon Raman spectrum for $E_{L}=2.0$ eV (bottom).
The constituting combination and overtone spectra are also shown (middle
and top).}
\end{figure}
\begin{figure}[tbph]
\includegraphics[width=80mm]{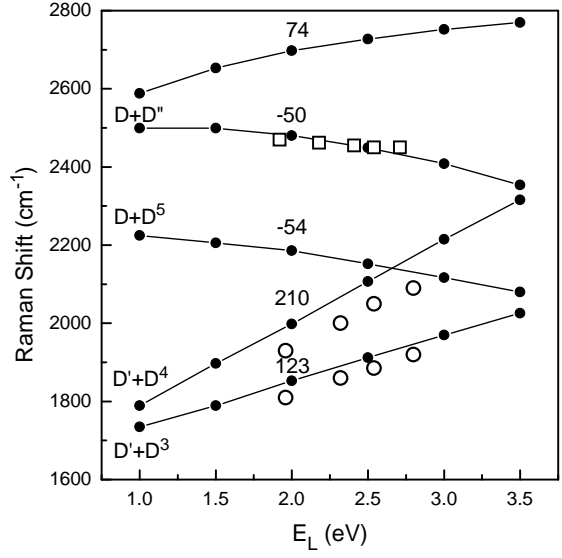} 
\caption{Calculated dependence of the Raman shift of the combination bands
on $E_{L}$ (solid symbols). The highest-frequency band is due to
TO$^{'}$LA$^{'}$@K phonons. The numbers are the slopes of the curves
for $E_{L}=2.0$ eV. The empty symbols - squares\cite{mafr07} and
circles,\cite{cong11} are experimental data.}
\end{figure}
\begin{figure}[tbph]
\includegraphics[width=80mm]{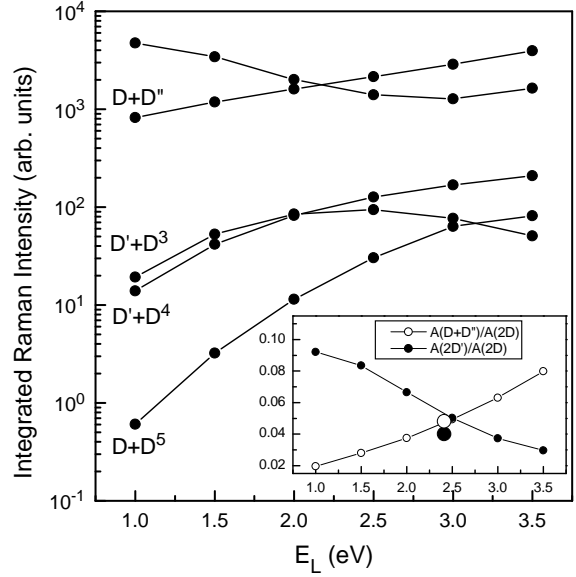} 
\caption{Calculated dependence of the integrated Raman intensity of the combination
bands on $E_{L}$ (solid symbols). The highest-intensity band at low
energies is due to TO$^{'}$LA$^{'}$@K phonons. Inset: The DFT-GW corrected ratios 
$A\left(2D^{'}\right)/A\left(2D\right)$ and $A\left(D+D^{''}\right)/A\left(2D\right)$ (small symbols) in comparison with experimental data\cite{ferr06} (large symbols). The scale is the same as in Fig. 7.}
\end{figure}
\begin{figure}[tbph]
\includegraphics[width=80mm]{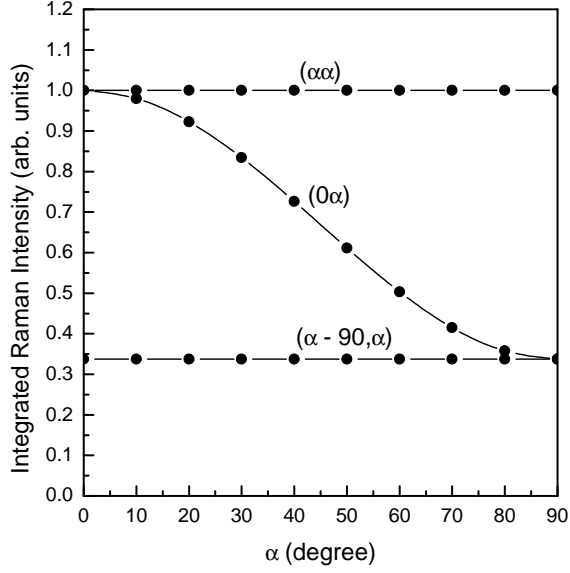} 
\caption{Calculated polarization dependence of the integrated Raman intensity
of the $2D$ band (solid symbols). The angle $\alpha$ is relative
the $x$ axis along a zigzag line of carbon bonds. The polarizer and
analyzer angles are given by the notation $\left(\alpha\beta\right)$.
The curves for parallel and cross polarization, $\left(\alpha\alpha\right)$
and$\left(\alpha-90,\alpha\right)$, as well for fixed polarizer direction
along the $x$ axis and variable analyzer orientation $\left(0\alpha\right)$
are shown. The solid line for polarization $\left(0\alpha\right)$
is a fit of Eq. (\ref{e8}). The obtained ratio of the integrated
intensities is $A_{\perp}/A_{\mid\mid}\approx0.34$.}
\end{figure}

\end{document}